\documentstyle[12pt,epsfig]{article}

\begin{document}
\topmargin -1.4cm
\oddsidemargin -0.8cm
\evensidemargin -0.8cm 

\title{Dimensional reduction in freely decaying\\ turbulent non-helical 
magnetic fields}

\vspace{1.5cm}

\author{P. Olesen\\
{\it  The Niels Bohr Institute}\\
{\it Blegdamsvej 17, Copenhagen \O, Denmark }}

\maketitle

\begin{abstract}
We show that the self-similarity property of decaying turbulent non-helical 
magnetic fields is the same in all dimensions. It is shown that these fields 
produce an inverse transfer in all dimensions. It is also shown that this 
phenomenon in a certain gauge can be 
assigned to a time independent value of the squared vector potential. This 
mechanism is similar to what happens in the well known 
inverse cascades in two dimensional
magnetohydrodynamics. 
\end{abstract}

\vskip0.5cm

Recently the non-helical\footnote{By non-helical we mean that
the magnetic helicity is small. Due to fluctuations of the fields helicity
can always have some small value $\neq 0$.} inverse transfer of a decaying 
turbulent magnetic
field has been studied in the papers \cite{axel} and \cite{zrake}. These
fields are for example of interest in cosmology and for optical polarization
in gamma ray burst afterglows. A 
few of the results obtained in these papers can be understood from the 
point of view of self-similarity of the energy, as discussed 
in our paper \cite{poul}. Self-similarity implies
 a motion  of the energy towards smaller $k$ as time goes on. In the following
we shall extend our previous study to show that the self-similarity is
independent of the dimension, and that there is an associated time independent
quantity, namely the square of the vector potential, as
found approximately numerically in the paper by   Kahniashvili, Brandenburg, 
and  Tevzadze   \cite{axel}. This is analogous to 
what happens in two dimensional MHD so for this reason one can say that a 
dimensional reduction occurs. 

We start from the following expression of the energy of the magnetic field
in $D$ dimension,
\begin{equation}
{\cal E}_B(k,t)=\frac{\Omega_Dk^{D-1}}{2(2\pi)^D}\int \frac{d^Dx}{V_D}
\int d^Dy~e^{i{\bf k(y-x)}}~
<{\bf B}({\bf x},t){\bf B}({\bf y},t)>,
\label{begin}
\end{equation}
where $\Omega_D$ is the full solid angle in $D$ dimensions and $V_D$ is a
$D$ dimensional normalization volume. The kinetic energy
${\cal E}_v(k,t)$ is the same expression with $\bf B$ replaced by
the velocity field $\bf v$. We assume that these fields are statistically 
isotropic. The normalization is given by
\begin{equation}
\int_0^\infty dk ~{\cal E}_B(k,t)=\frac{1}{2}~\int\frac{d^Dx}{V_D}
<{\bf B}({\bf x},t)^2 >
=\frac{1}{2}~<{\bf B}({\bf x},t)^2 >.
\end{equation}

Proceeding as in \cite{poul} by use of the following scaling invariance of
the MHD equations 
\begin{equation}
{\bf x}\rightarrow l{\bf x},~t\rightarrow l^2t,~{\bf v}\rightarrow {\bf v}/l,
{\bf B}\rightarrow {\bf B}/l,~\nu\rightarrow \nu,~\eta\rightarrow \eta,~p
\rightarrow p/l^2,
\label{scale}
\end{equation}
where $\nu$ and $\eta$ are the kinetic and Ohmic diffusions, respectively, we
obtain from Eq. (\ref{begin})
according to these scalings (with ${\bf x}=l{\bf x'},
{\bf y}=l{\bf y'}$, and $V_D=l^DV_D'$)
\begin{equation}
{\cal  E}_B (k/l,l^2t)=l~\frac{\Omega_D k^{D-1}}{2(2\pi)^D}~
\int\frac{d^Dx'}{V_D'}
\int d^Dy'~
e^{i{\bf k (y'-x')}}< {\bf B}(l{\bf x'},l^2t){\bf B}(l{\bf y'},l^2t)>
=l^{-1}{\cal E}_B(k,t),
\label{r}
\end{equation}
where ${\bf x}=l{\bf x'},{\bf y}=l{\bf y'}$, and $V_D=l^DV_D'$. There is a 
similar expression for the kinetic energy
with $\bf B$ replaced by the velocity field $\bf v$. 

As in \cite{poul} we obtain the self-similarity (choose $l=\sqrt{t_0/t}$)
\begin{equation}
{\cal E}_{v,B}(k,t)=\sqrt{t_0/t}~{\cal E}_{v,B}
\left(k\sqrt{t/t_0},t_0\right),
\label{stuff}
\end{equation}

The important point is that this relation  is {\it independent~of~
 the dimension D}. This is due in part to the dimensional indifference of 
the scaling relations (\ref{scale}) and in part by the way the dimension 
enters Eq. (\ref{r}). Therefore all results derived from the
self-similarity are independent of the dimension.

The scaling arguments presented above imply the relation (\ref{stuff}), but
does not tell us anything about the energies as functions of $k$. These
functions could quite well depend on the dimensions. Therefore, in the
following it should be kept in mind that it is only the self-similarity which 
is indifferent to the dimensions.

As shown in \cite{poul} self-similarity leads to some quantitative results,
\begin{equation}
{\cal E}_{v,B}(t)=\int_0^\infty dk~{\cal E}_{v,B}(k,t)=
\frac{t_0}{t}\int_0^\infty dx~{\cal E}_{v,B}(x,t_0)\propto \frac{1}{t},~
{\rm all~dimensions},
\end{equation}
and
\begin{equation}
<\xi>=\kappa^{-1}(t)=\int_0^\infty~\frac{dk}{ k}~{\cal E}(k,t)/
 {\cal E}(t) =\sqrt\frac{t_0}{t}~\int_0^\infty\frac{dx}{x}
{\cal E}(x,t_0)/{\cal E}(t)\propto \sqrt{t},~ {\rm all~dimensions}.
\end{equation}
For three dimensions these results are in accordance with the numerical 
calculations reported in \cite{axel}.

Although we want to discuss non-helical magnetic fields there are always
fluctuations \cite{axel} leading to some small magnetic helicity as
pointed out in \cite{axel}. In \cite{poul}
we showed that the integrated helicity decreases like $1/\sqrt{t}$, i.e.
slower than the energy\footnote{For a discussion of the combination of
scaling and helicity
we refer to the paper by Campanelli \cite{campanelli}.}

The transfer of energy from smaller to larger scales can be understood as a 
generic feature of freely decaying turbulence in magnetohydrodynamics through 
the self-similarity (\ref{stuff}), which in turn is a consequence of the 
dimensionally indifferent scaling (\ref{scale}). In spite of this it may be
of interest to study whether the inverse behavior can be understood in
more conventional terms. In the paper \cite{axel} by  Kahniashvili, 
Brandenburg, and Tevzadze there is a discussion (see p. 8 and p. 11) of the
inverse cascade phenomenon. In three dimensions this is related to 
the magnetic helicity,
whereas in two dimensions it is related to $<A^2>$, where $A$ is the vector
 potential in two dimensions. 
In the paper \cite{axel} the authors speculate 
whether the inverse transfer observed in their calculations could be
due to a (locally) two dimensional flow occurring in three dimensions.
They also find numerically that $<A^2>$ is approximately constant in time.
In the following we shall investigate this idea from the point of 
self-similarity.

 The vector potential is defined through $\bf B=\nabla\times A$. Obviously 
$<A^2>$  is gauge dependent. We choose the Lorenz gauge
\begin{equation}
\partial_i A_i=0,~~A_i\rightarrow \tilde{A}_i+\partial_i\Lambda,~{\rm and}~
\partial^2\Lambda=0.
\label{gauge}
\end{equation}
The scaling property of $\bf A$ follows from Eq. (\ref{scale}), according to
which
${\bf B}\rightarrow {\bf B}/l$, so $\bf A $ remains invariant,
\begin{equation}
{\bf A}({\bf x},t)\rightarrow {\bf A}(l{\bf x},l^2t).
\label{s}
\end{equation}
Next consider the quantity
\begin{equation}
{\cal AA}(k,t)=\frac{\Omega_Dk^{D-1}}{(2\pi)^D}\int\frac{d^Dx}{V}
\int d^Dy ~e^{i\bf k(y-x)}
<{\bf A}({\bf x},t){\bf A}({\bf y},t)>,
\label{newnew}
\end{equation}
with the property
\begin{equation}
\int dk {\cal AA}(k,t)=\int\frac{d^Dx}{V}<{\bf A}({\bf x},t)^2>=
<{\bf A}({\bf x},t)^2>
\label{norm}
\end{equation}
The expression (\ref{newnew}) is invariant under the gauge transformations
(\ref{gauge}), as is seen by some partial integrations in (\ref{newnew}).
First, the factor in the integrand
\begin{equation}
<A_i({\bf x},t)A_i(({\bf y},t)>
\end{equation}
by a gauge transformation of the type (\ref{gauge}) becomes
\begin{eqnarray}
<\tilde{A}_i({\bf x},t)\tilde{A}_i({\bf y},t)>+<\tilde{A}_i({\bf x},t)
\partial_i\Lambda({\bf y},t)>+<\partial_i\Lambda({\bf x},t)
\tilde{A}_i({\bf y},t)> \\ \nonumber
+<\partial_i \Lambda ({\bf x},t)\partial_i\Lambda({\bf y},t)>.
\label{xxx}
\end{eqnarray}
Then by  partial integrations e.g. the $\partial_y$ differentiations can be 
shifted to the exponent in (\ref{newnew}) and replaced by $-\partial_x$
because $\bf x-y$ enters the exponent.
 Another partial integration  then shifts the differentiations back to the 
$<>$ brackets\footnote{It is assumed that the contributions from the boundaries
vanish, e.g. because the fields vanish at infinity or by suitable periodic
boundary conditions.}, so
the right hand side becomes
\begin{eqnarray}
 <\tilde{A}_i({\bf x},t)\tilde{A}_i({\bf y},t)>-<\partial_i\tilde{A}_i
({\bf x},t)\Lambda({\bf y},t)>-<\Lambda({\bf x},t)\partial_i
\tilde{A}_i({\bf y},t)>\\ \nonumber 
-<\partial^2 \Lambda ({\bf x},t)\Lambda({\bf y},t)>.
\end{eqnarray}
By use of the gauge conditions (\ref{gauge}) we see that the last
three terms vanish. Thus we have shown that
\begin{equation}
\int\frac{d^Dx}{V}
\int d^Dy ~e^{i\bf k(y-x)}<A_i({\bf x},t)A_i(({\bf y},t)>=
\int\frac{d^Dx}{V}
\int d^Dy ~e^{i\bf k(y-x)}<\tilde{A}_i({\bf x},t)\tilde{A}_i(({\bf y},t)>
\end{equation}
So the expression
(\ref{newnew}) is invariant under the gauge transformation (\ref{gauge}).

Using the scaling (\ref{s}) in Eq. (\ref{newnew}) we obtain (with 
${\bf x}=l{\bf x'},{\bf y}=l{\bf y'}$, and $V_D=l^DV_D'$)
\begin{equation}
{\cal AA}(k/l,l^2t)=l\frac{\Omega_Dk^{D-1}}{(2\pi)^D}\int\frac{d^Dx'}{V'}
\int d^Dy' ~e^{i\bf k(y'-x')}
<{\bf A}(l{\bf x'},l^2t){\bf A}(l{\bf y}',l^2t)>=l{\cal AA}(k,t).
\label{1}
\end{equation}
This relation has to be valid for all $l$, so taking as before 
\begin{equation}
l=\sqrt{t_0/t}
\end{equation}
Eq. (\ref{1}) gives
\begin{equation}
{\cal AA}(k,t)=\sqrt{t/t_0}~{\cal AA}(k\sqrt{t/t_0},t_0).
\end{equation}
Inserting this self-similarity relation in Eq.(\ref{norm}) we obtain
\begin{equation}
<{\bf A}({\bf x},t)^2>=\int dk {\cal AA}(k,t)=\int dx {\cal AA}(x,t_0)=
{\rm independent~ of~ }t,
\end{equation}
where $x=k\sqrt{t/t_0}$.

We have therefore arrived at the result that in the gauge (\ref{gauge})
the quantity $<A^2>$ is constant, like in two 
dimensional MHD, This conserved quantity is obviously not in general
gauge invariant,
but the scaling leading to time independence is valid at least in the 
particular gauge (\ref{gauge}).

In conclusion we have shown that the dimensional reduction based on
$<A^2>$=constant found approximately
in the paper \cite{axel} is valid exactly in the gauge (\ref{gauge}) using
simple scaling arguments. The constancy of $<A^2>$ 
is thus a consequence of
self-similarity \cite{poul}, which we have shown to be valid in the 
same form in all dimensions.


\begin{thebibliography}{X}
\bibitem{axel}T. Kahniashvili, A. Brandenburg, and A. Tevzadze, {\it Nonhelical
inverse transfer of a decaying turbulent magnetic field},
arXiv:1404.2238; Phys. Rev. Lett. 114 (2015) 7,075001.
\bibitem{zrake}J. Zrake, {\it Inverse cascade of non-helical magnetic 
turbulence in 
a relativistic fluid}, arXiv:1407.5626; Astrophys. J. 809 (2015) 1,39.
\bibitem{poul}P. Olesen, {\it Inverse transfer of self-similar decaying 
turbulent non-helical magnetic field}, arXiv:1509.08962.
\bibitem{campanelli}L. Campanelli, {\it Scaling laws in magnetohydrodynamic 
turbulence}, Phys. Rev. D70 (2004) 083009.
\end{thebibliography}
\end{document}